# Applying Deep Learning to Specific Learning Disorder Screening

**Nuriel S. Mor, Ph.D.**
Licensed psychologist
Software Engineering Program with an Emphasis on AI
Darca and Bnei Akiva Schools, Israel.
Talpiot College of Education, Israel.

**Kathryn L. Dardeck, Ed.D.**
Licensed psychologist, Health Service Provider
Adjunct faculty, William James College, Massachusetts, USA.

## Abstract

Early detection is key for treating those diagnosed with specific learning disorder, which includes problems with spelling, grammar, punctuation, clarity and organization of written expression. Intervening early can prevent potential negative consequences from this disorder. Deep convolutional neural networks (CNNs) perform better than human beings in many visual tasks such as making a medical diagnosis from visual data. The purpose of this study was to evaluate the ability of a deep CNN to detect students with a diagnosis of specific learning disorder from their handwriting. The MobileNetV2 deep CNN architecture was used by applying transfer learning. The model was trained using a data set of 497 images of handwriting samples from students with a diagnosis of specific learning disorder, as well as those without this diagnosis. The detection of a specific learning disorder yielded on the validation set a mean area under the receiver operating characteristics curve of 0.89. This is a novel attempt to detect students with the diagnosis of specific learning disorder using deep learning. Such a system as was built for this study, may potentially provide fast initial screening of students who may meet the criteria for a diagnosis of specific learning disorder.

**Keywords:** *specific learning disorder, deep learning, deep CNNs, transfer learning*

*We wish to thank teaching assistant Karin Volovik for her assistance in gathering and processing data for this study.*

*Please address queries to: Nuriel S. Mor. Email: nuriel.mor@gmail.com*

## Introduction

Specific learning disorder is a neurodevelopmental disorder that can be detected only after formal education starts (American Psychiatric Association, 2013). About 10 percent of school-age children are diagnosed as having this disorder (Fortes et al., 2016; Gorker et al., 2017). Specific learning disorder can manifest in several different academic areas including reading, writing and mathematics (American Psychiatric Association, 2013). When this diagnosis is about an impairment in reading, symptoms may include difficulty with word accuracy, reading fluency, and reading comprehension. In impairment in written expression, symptoms may include difficulty with spelling, grammar, punctuation and organization. Mathematical impairments may include memorization of mathematical facts, fluent calculation and mathematical reasoning (American Psychiatric Association, 2013). The aforementioned symptoms are further clarified according to severity of mild, moderate or severe (American Psychiatric Association, 2013). A diagnosis of specific learning disorder is complex and made through a combination of observation, interviews, family history, and school reports (American Psychiatric Association, 2013; McDonough et al., 2017).

Early detection is vital for children with specific learning disorder. If this diagnosis is undetected, detrimental consequences including high levels of psychological distress, depression, suicidality, and poorer overall mental health may ensue (American Psychiatric Association, 2013). On the other hand, early detection and intervention can significantly mitigate the negative impact of specific learning disorder on mental health (American Psychiatric Association, 2013). Early diagnosis helps in preventing the frustration and decrease in wellbeing caused by an undiagnosed specific learning disorder (Lombardi et al., 2019).

**Deep Learning and Diagnosis**

Deep learning algorithms are more accurate than human beings in many visual tasks such as strategic board games, human and chimpanzee facial recognition, plant disease identification, and object recognition (Esteva et al., 2017; Ferentinos et al., 2018; Schofield et al., 2019). In addition, deep learning algorithms perform better than humans in medical diagnosis based on visual data such as skin cancer classification, breast cancer screening, and pneumonia detection (Esteva et al., 2017; McKinney et al.,2020; Rajpurkar et al., 2017). Advances in computation, very large datasets and emerging new techniques enable deep learning algorithms to recognize very complex patterns in data that are beyond human perception (Esteva et al., 2017).

The medical diagnostic world is fundamentally affected by this progress as we witness more and more successful deep learning applications that help with the medical diagnostic process (Esteva et al., 2017; Kermany et al., 2018; McKinney et al ., 2020; Rajpurkar et al., 2017). Deep learning applications for mental disorder screening have been based mainly on data from neuroimaging (Galatzer-Levy, Karstoft, Statnikov, & Shalev, 2014; Vieira, Pinaya, & Mechelli, 2017). A range of psychiatric and neurological disorders such as post-traumatic stress disorder, depression, schizophrenia and more, can be screened from neuroimaging data using deep learning (Vieira et al., 2017). In addition, neurodevelopmental disorders such as attention deficit hyperactivity disorder and autism spectrum disorder can be screened from neuroimaging data with deep learning (Heinsfeld, Franco, Craddock, Buchweitz, & Meneguzzi, 2018; Vieira et al., 2017).

Only a few studies (Gurovich et al., 2019; Mor & Dardeck, 2018; Rad et al., 2018; Shukla, Gupta, Saini, Singh, & Balasubramanian, 2017) have been published on using deep learning that do not employ neuroimaging to flag possible mental disorders. This fact impedes the implementation of deep learning in the diagnostic screening process of mental disorders because neuroimaging is rarely used in psychology because of its high cost (Galatzer-Levy et al., 2014).

Mor and Dardeck (2018) succeeded in identifying people at risk for Post Traumatic Stress Disorder (PTSD) using readily collected ecological risk factors and deep learning. Shukla et al. (2017) succeeded in detecting developmental disorders from facial images using deep learning. They succeeded in building a deep learning model that performs better than humans in recognizing and differentiating among a spectrum of neurodevelopmental disorders including autism spectrum disorder, fetal alcohol syndrome, Down syndrome, progeria, cerebral palsy, and intellectual disability. In addition, other researchers (Gurovich et al., 2019) built a deep learning model that identifies facial phenotypes of more than 200 genetic syndromes such as Lubs XL MR, fragile X MR, Prader–Willi, MR XL Bain type, Angelman, Ch1p36 del, fetal alcohol, Potocki–Lupski, Rett and many more. Rad et al., (2018) successfully detected stereotypical motor movement in patients with autism spectrum disorder with the aid of deep learning.

**Deep Learning and Specific Learning Disorder Diagnosis**

Specific learning disorder may affect handwriting in a way that can be visually distinguished (Li-Tsang, Lau, Ho, & Leung, 2018). Symptoms of specific leaning disorder may include impairment in written expression such as difficulty with spelling, grammar, punctuation, and organization (American Psychiatric Association, 2013). Handwriting performance and sensorimotor skills may be affected by specific leaning disorder (Li-Tsang et al., 2018). Students with specific learning disorder write at slower speed, and with greater variation in written character size (Lam, Au, Leung , & Li-Tsang, 2011). They require more time to fulfill handwriting assignments in class (Engel-Yeger, Nagauker-Yanuv, & Rosenblum, 2009). Engel-Yeger et al., (2009) suggested that their movements were less mature than non-learning disordered students, and their performance less accurate in space and time. Students with the diagnosis of specific learning disorder were found to erase more and complain about fatigue (2009). The legibility of their handwriting was found to be poor compared to handwriting of students without this disorder (2009).

The purpose of the current study was to evaluate the ability of deep learning to distinguish between those who have a specific learning disorder and those who do not, from their handwriting. Outfitted with deep learning, mobile devices can assist with the rapid screening of students with specific learning disorder based on their handwriting. This in turn, may contribute to early detection and intervention after a careful follow-up evaluation.

## Method

**Sample and Outcome Measure**

The target population for this study included high school students between 15 and 18 years old from Hadash High School, Bat-Yam, Israel. Handwriting samples were collected from 152 students who volunteered to participate in this study. No remuneration was promised or given. Students volunteered to provide their old notebooks. About 500 pages of handwriting were scanned and saved as images. Two completely sealed and locked boxes were placed in one classroom, for a few hours after the school day, for 2 consecutive days. One box was intended for notebooks of students who had been previously diagnosed as having specific learning disorder, while the other was designed for students without specific learning disorder. Diagnosing the students had previously been done and was unrelated to this study. The notebook collection process was voluntarily conducted with complete anonymity. The outcome measure of this study is a dichotomized variable of no diagnosis of specific learning disorder versus diagnosis of specific learning disorder.

**Modeling Approach**

Deep convolutional neural networks (CNNs) are the state of the art technology in visual tasks (Esteva et al., 2017). MobileNetV2 is a deep CNN which achieves cutting edge results in visual tasks (Sandler, Howard, Zhu, Zhmoginov, & Chen, 2018). The great benefit of MobileNet models is that they were designed to be deployed on mobile devices, allowing a rapid inference from a photo taken on a mobile device (Howard et al., 2017; Sandler et al., 2018). MobileNet models were trained on the ImageNet dataset which contains more than 14 million images with 1000 object categories (Howard et al., 2017; Sandler et al., 2018). MobileNet models specialize and excel in several visual tasks including object detection, face attributes, fine-grain classification, and landmark recognition (Howard et al., 2017), as demonstrated in figure 1.

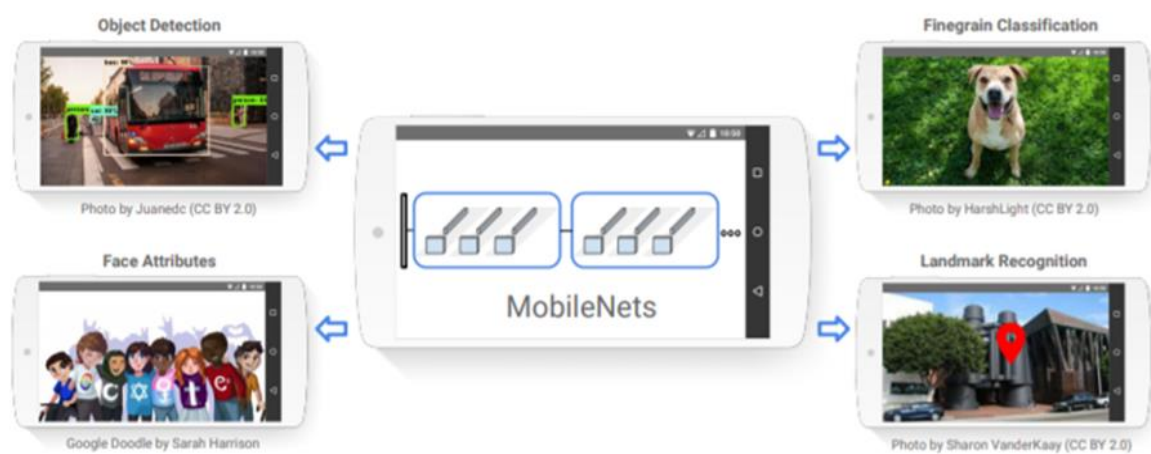

**Figure 1.** MobileNet models can be applied to various recognition tasks for efficient on device intelligence. From Howard et al., 2017.

Transfer learning is a technique where a model developed for a task is reused as the starting point for a model on a second task. This technique involves removing the last layer of the pre-trained deep neural network, adding new layers suitable for a current specific task, and training with a new dataset (Esteva et al., 2017; Khan et al., 2019). Transfer learning is a very useful technique in which researchers can utilize pre-trained, state of the art deep neural networks (Khan, Islam, Jan, Din, & Rodrigues, 2019).

In this study, the pre-trained MobileNetV2 (Sandler et al., 2018) architecture was utilized using transfer learning. MobileNetV2 is a suitable architecture for transfer learning in visual tasks as needed in this study. The last SoftMax layer of the MobileNetV2 architecture designed for classification of 1000 different classes of the ImageNet dataset was removed, and 3 hidden layers of Relu neurons were added: layer 1 of 800 neurons, layer 2 of 400 neurons, and layer 3 of 200 neurons. Additionally, the last layer of a single sigmoid neuron for classifying the 2 desired classes in this study was added: no diagnosis of specific learning disorder versus diagnosis of specific learning disorder. Table 1 presents the deep neural network architecture and model summary using MobileNetV2 and transfer learning. In addition, dropout and data augmentation were used, two techniques that enhance the performance and generalizability of deep neural networks (Perez & Wang, 2017; Srivastava, Hinton, Krizhevsky, Sutskever, & Salakhutdinov, 2014).

**Table 1:** *Model Summary*

| Layer | Output Shape | Param # |
| --- | --- | --- |
| Keras Layer | (None, 1280) | 2257984 |
| Dense | (None, 800) | 1024800 |
| Dropout | (None, 800) | 0 |
| Dense | (None, 400) | 320400 |
| Dropout | (None, 400) | 0 |
| Dense | (None, 200) | 80200 |
| Dropout | (None, 200) | 0 |

Total params: 3,683,585
Trainable params: 1,425,601
Non-trainable params: 2,257,984

**Validation and Accuracy Metrics**

The collected data set of 497 images of handwriting was randomly split to a training set containing 447 images, and a validation set containing 50 images. Five metrics were used to estimate the performance of the deep learning model: area under the curve (AUC), precision, recall, F-score, and accuracy. All the metric values reported in this study, represent results obtained from the validation set.

Area under the curve is the area between the receiver operating characteristic (ROC) curve and the x-axis. The receiver operating characteristic curve is defined by plotting the true positive rate against the false-positive rate at different thresholds (Majnik & Bosnić, 2013). The area between the receiver operating characteristic is an unbiased metric of performance and can be compared to AUC of different systems (Karstoft, Statnikov, Andersen, Madsen, & Galatzer-Levy, 2015). Precision is defined by true positives divided by the sum of true positives and false positives (Goutte & Gaussier, 2005). Recall is defined by true positives divided by the sum of true positives and false negatives (Goutte & Gaussier, 2005). The F-score is a balanced metric, defined by a weighted average of precision and recall (Hand & Christen, 2018). Accuracy is defined by all true predictions of the model divided by the total of all predictions (Sim et al., 2019).

## Experiments and Results

**Descriptive Statistics**

All the students who provided their notebooks were high school students from Hadash High School, Bat-Yam, Israel. They were all between 15 and 18 years old. Consistent with the prevalence of specific learning disorder reported in the literature (American Psychiatric Association, 2013), 17 of the 152 students who participated (11%) had the diagnosis of specific learning disorder.

**Main Analyses**

The model was trained for 25 epochs. The model yielded the best accuracy after 21 epochs and started to decline from epoch 22, as expected because of overfitting (Cha et al., 2019). Figure 2 shows the working system. The model yielding the best accuracy was saved for further analysis of performance metrics. The model yielded: AUC= 0.89, precision=0.94, recall =0.89, F-score=0.91, and accuracy=0.92. Figure 3 presents the changes in accuracy during training.

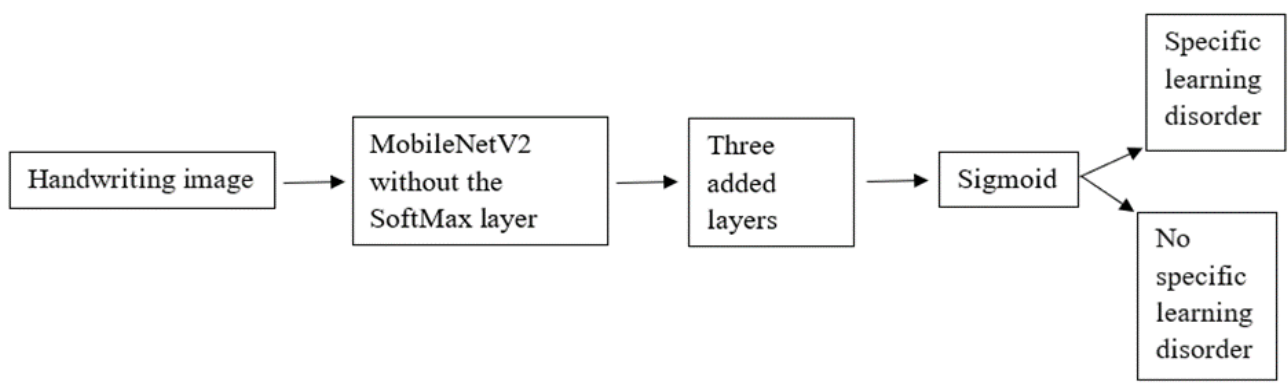

**Figure 2.** *The working system.*

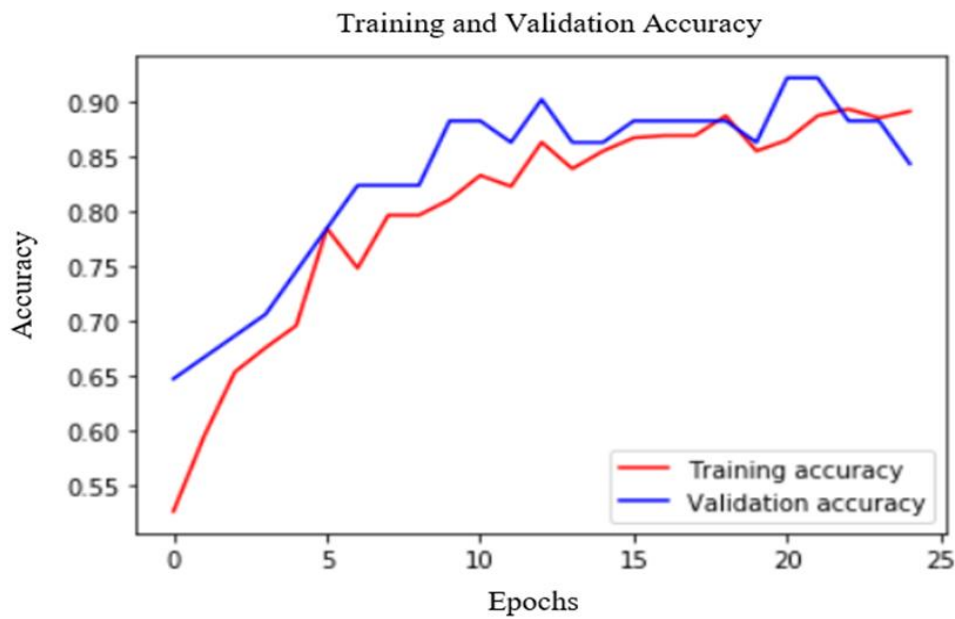

**Figure 3.** *Accuracy during training: the model yielded the best accuracy after 21 epochs.*

## Discussion

This study evaluated the ability of deep learning algorithms to screen students with specific learning disorder by using their handwriting. This was the first study that applied deep learning to screening for specific learning disorder classification from handwriting samples that were easily collected for fast inference and detection.

The present study model yielded an AUC of 0.89. This indicates a good predictive model in the domain of mental diagnostics (Galatzer-Levy et al., 2014). Precision of 0.94, recall of 0.89, F-score of 0.91, and accuracy of 0.92 indicate that the model yields very good results in specific learning disorder detection, compared to other studies of mental disorder detection with deep learning (Vieira et al., 2017). Values of performance metrics in other studies using deep learning to detect mental disorders from neuroimaging data were between 0.65 to 0.95 (Vieira et al., 2017). The reported accuracy of the model designed to identify facial phenotypes of genetic disorders using deep learning was 0.91 (Gurovich et al., 2019). The AUC and F-score of the model designed to identify people at risk for PTSD using ecological factors and deep learning were 0.91 and 0.83 respectively (Mor & Dardeck, 2018).

The finding that deep learning applied to handwriting samples provides efficient initial screening of students for specific learning disorder is promising. About 10% of school-age children have specific learning disorder (American Psychiatric Association, 2013; Fortes et al., 2016). Screening of specific learning disorder using handwriting and deep learning can make the complex task of specific learning disorder diagnosis faster and simpler.

It is important to mention that we are not suggesting that such a model would replace the essential diagnostic process in which mental health professionals consider a combination of information from observations, interviews, family history, and school reports (American Psychiatric Association, 2013). We are suggesting, however, that a model such as the one designed for this study can provide fast-initial screening of students for specific learning disorder. This could, therefore, significantly contribute to early detection and intervention.

**Applicability of This Study**

About 6 billion smartphone subscriptions will exist by the end of 2020 (Esteva et al., 2017). Smartphone applications that can help with the initial screening of medical or mental disorders would provide low-cost universal access to essential diagnostic care (Esteva et al., 2017). The deep learning model built in this study is based on MobileNet which was designed for smartphones (Howard et al., 2017). MobileNet provides fast and accurate performance deployed on mobile devices (Howard et al., 2017). Outfitted with a CNN, mobile devices can aid educators, reading specialists, and other relevant professionals with a means to achieve fast initial screening of specific learning disorder. Screening for students with specific learning disorder using this system, requires no more than taking a photo of handwriting on a smartphone, uploading, and sending it to the model, and receiving the model answer. For further edification, the system designed in this study may be viewed at

https://colab.research.google.com/drive/1SUByhCjS29pR_njEwFKD7v3YFZ9C_i9H

**Limitations and Recommendation for Future Work**

This study was conducted on students of Hadash High School, Bat Yam, Isarel. The results of this initial study cannot be generalized beyond this specific Hebrew speaking population. It would be important and interesting to assess handwriting of students using multiple languages to get a picture as to how the algorithm holds up across different alphabets and writing systems. In order to increase the generalizability of our model, the main recommendations for future work include collecting handwriting samples from many different populations in many different languages, thereby significantly increasing the size of the handwriting data set. The size of the training data set is the most important factor for enhancing the generalizability of deep learning models (Perez & Wang, 2017).

**Summary, Conclusion and Future Directions**

This study demonstrated the feasibility of screening students with specific learning disorder from handwriting using a deep learning algorithm. The model designed in this study can be easily deployed on smartphones, enabling fast initial screening of students with specific learning disorder simply by taking a photo of their handwriting. Early intervention is essential for children with specific learning disorder, and such a system as developed in this study may significantly contribute to early detection, and subsequent intervention. The system in this study is far from a universal, optimal solution because the training data set was limited. It is hoped, however, that the study's findings will serve as an inspiration for the future development of a universal solution for early screening and detection of specific learning disorder, which would ideally include many different populations from across the world.

## References

American Psychiatric Association. (2013). Diagnostic and statistical manual of mental disorders (5th ed.). Arlington, VA: Author

Cha, K. H., Petrick, N., Pezeshk, A., Graff, C. G., Sharma, D., Badal, A., ... & Sahiner, B. (2019, March). Reducing overfitting of a deep learning breast mass detection algorithm in mammography using synthetic images. In *Medical Imaging 2019: Computer-Aided Diagnosis* (Vol. 10950, p. 1095004). International Society for Optics and Photonics.

Engel-Yeger, B., Nagauker-Yanuv, L., & Rosenblum, S. (2009). Handwriting performance, self-reports, and perceived self-efficacy among children with dysgraphia. *American Journal of Occupational Therapy*, *63*(2), 182-192.

Esteva, A., Kuprel, B., Novoa, R. A., Ko, J., Swetter, S. M., Blau, H. M., & Thrun, S. (2017). Dermatologist-level classification of skin cancer with deep neural networks. *Nature*, *542*(7639), 115.

Ferentinos, K. P. (2018). Deep learning models for plant disease detection and diagnosis. *Computers and Electronics in Agriculture*, *145*, 311-318.

Fortes, I. S., Paula, C. S., Oliveira, M. C., Bordin, I. A., de Jesus Mari, J., & Rohde, L. A. (2016). A cross-sectional study to assess the prevalence of DSM-5 specific learning disorders in representative school samples from the second to sixth grade in Brazil. European child & adolescent psychiatry, 25(2), 195-207.

Galatzer-Levy, I. R., Karstoft, K. I., Statnikov, A., & Shalev, A. Y. (2014). Quantitative forecasting of PTSD from early trauma responses: A machine learning application. Journal of Psychiatric Research, 59, 68–76

Gorker, I., Bozatli, L., Korkmazlar, Ü., Karadag, M. Y., ceylan, C., Sogut, C., ... & Turan, N. (2017). The probable prevalence and sociodemographic characteristics of specific learning disorder in primary school children in Edirne. Archives of Neuropsychiatry, 54(4), 343.

Goutte, C., & Gaussier, E. (2005, March). A probabilistic interpretation of precision, recall and F-score, with implication for evaluation. In *European Conference on Information Retrieval* (pp. 345-359). Springer, Berlin, Heidelberg.

Gurovich, Y., Hanani, Y., Bar, O., Nadav, G., Fleischer, N., Gelbman, D., ... & Bird, L. M. (2019). Identifying facial phenotypes of genetic disorders using deep learning. Nature medicine, 25(1), 60.

Hand, D., & Christen, P. (2018). A note on using the F-measure for evaluating record linkage algorithms. Statistics and Computing, 28, 539–547

Heinsfeld, A. S., Franco, A. R., Craddock, R. C., Buchweitz, A., & Meneguzzi, F. (2018). Identification of autism spectrum disorder using deep learning and the ABIDE dataset. *NeuroImage: Clinical*, *17*, 16-23.

Howard, A. G., Zhu, M., Chen, B., Kalenichenko, D., Wang, W., Weyand, T., ... & Adam, H. (2017). Mobilenets: Efficient convolutional neural networks for mobile vision applications. arXiv preprint arXiv:1704.04861.

Lam, S. S., Au, R. K., Leung, H. W., & Li-Tsang, C. W. (2011). Chinese handwriting performance of primary school children with dyslexia. *Research in developmental disabilities*, *32*(5), 1745-1756.

Li-Tsang, C. W., Li, T. M., Lau, M. S., Ho, C. H., & Leung, H. W. (2018). Handwriting assessment to distinguish comorbid learning difficulties from attention deficit hyperactivity disorder in C hinese adolescents: A case–control study. *International journal of methods in psychiatric research, 27(4), e1718.*

Lombardi, E., Offredi, I., Bettoni, R., Sarti, D., Traficante, D., & Vernice, M. (2019). Well-being in secondary school students: a comparison between specific learning disorder and difficulties. In *1st SRLD Conference* (pp. 114-115). Cleup.

Karstoft, K. I., Statnikov, A., Andersen, S. B., Madsen, T., & Galatzer-Levy, I. R. (2015). Early identification of posttraumatic stress following military deployment: Application of machine learning methods to a prospective study of Danish soldiers. *Journal of affective disorders, 184, 170–175.*

Khan, S., Islam, N., Jan, Z., Din, I. U., & Rodrigues, J. J. C. (2019). A novel deep learning based framework for the detection and classification of breast cancer using transfer learning. Pattern Recognition Letters, 125, 1-6.

Kermany, D. S., Goldbaum, M., Cai, W., Valentim, C. C., Liang, H., Baxter, S. L., ... & Dong, J. (2018). Identifying medical diagnoses and treatable diseases by image-based deep learning. *Cell*, *172*(5), 1122-1131.

Majnik, M., & Bosnić, Z. (2013). ROC analysis of classifiers in machine learning: A survey. Intelligent Data Analysis, 17, 531–558.

McKinney, S. M., Sieniek, M., Godbole, V., Godwin, J., Antropova, N., Ashrafian, H., ... & Etemadi, M. (2020). International evaluation of an AI system for breast cancer screening. *Nature*, *577*(7788), 89-94.

McDonough, E. M., Flanagan, D. P., Sy, M., & Alfonso, V. C. (2017). Specific learning disorder. In *Handbook of DSM-5 Disorders in Children and Adolescents* (pp. 77-104). Springer, Cham.

Mor, N. S., & Dardeck, K. L. (2018). Quantitative Forecasting of Risk for PTSD Using Ecological Factors: A Deep Learning Application. *Journal of Social, Behavioral, and Health Sciences*, *12*(1), 4.

Perez, L., & Wang, J. (2017). The effectiveness of data augmentation in image classification using deep learning. *arXiv preprint arXiv:1712.04621*.

Powers, D. M. (2011). Evaluation: from precision, recall and F-measure to ROC,

Rad, N. M., Kia, S. M., Zarbo, C., van Laarhoven, T., Jurman, G., Venuti, P., ... & Furlanello, C. (2018). Deep learning for automatic stereotypical motor movement detection using wearable sensors in autism spectrum disorders. *Signal Processing*, *144*, 180-191.

Rajpurkar, P., Irvin, J., Zhu, K., Yang, B., Mehta, H., Duan, T., ... & Lungren, M. P. (2017). Chexnet: Radiologist-level pneumonia detection on chest x-rays with deep learning. *arXiv preprint arXiv:1711.05225*.

Sandler, M., Howard, A., Zhu, M., Zhmoginov, A., & Chen, L. C. (2018). Mobilenetv2: Inverted residuals and linear bottlenecks. In *Proceedings of the IEEE Conference on Computer Vision and Pattern Recognition* (pp. 4510-4520).

Schofield, D., Nagrani, A., Zisserman, A., Hayashi, M., Matsuzawa, T., Biro, D., & Carvalho, S. (2019). Chimpanzee face recognition from videos in the wild using deep learning. Science Advances, 5(9), eaaw0736.

Shukla, P., Gupta, T., Saini, A., Singh, P., & Balasubramanian, R. (2017, March). A deep learning frame-work for recognizing developmental disorders. In 2017 IEEE Winter Conference on Applications of Computer Vision (WACV) (pp. 705-714). IEEE.

Sim, J. Y., Jang, Y., Kim, W. C., Kim, H. Y., Lee, D. H., & Kim, J. H. (2019). Comparing the accuracy (trueness and precision) of models of fixed dental prostheses fabricated by digital and conventional workflows. Journal of prosthodontic research, 63(1), 25-30.

Srivastava, N., Hinton, G., Krizhevsky, A., Sutskever, I., & Salakhutdinov, R. (2014). Dropout: A simple way to prevent neural networks from overfitting. The Journal of Machine Learning Research, 15, 1929–1958.

Vieira, S., Pinaya, W. H., & Mechelli, A. (2017). Using deep learning to investigate the neuroimaging correlates of psychiatric and neurological disorders: Methods and applications. *Neuroscience & Biobehavioral Reviews*, *74*, 58-75.